\documentstyle[sprocl,epsfig]{article}

\def\be{\begin{equation}}
\def\ee{\end{equation}}
\def\bea{\begin{eqnarray}}
\def\eea{\end{eqnarray}}
\begin{document}

  \title{STUDY OF DARK MATTER INSPIRED cMSSM SCENARIOS\\ 
AT A TeV-CLASS LINEAR COLLIDER\\}

\author{MARCO BATTAGLIA}

\address{Department of Physics\\
University of California, Berkeley\\
and Lawrence Berkeley National Laboratory\\
Berkeley, CA 94720, USA}


\maketitle\abstracts{}

%***********************************************************************
\section{Introduction}
%***********************************************************************

The connections between Cosmology and Particle Physics are receiving 
significant attention for sharpening the physics case for a Linear Collider 
(LC), operating at energies up to about 1~TeV. Not only the anticipated 
accuracy of the LC data appears best suited for a deeper understanding of 
the cryptic messages obtained by satellite experiments. Cosmology data 
also offer an important opportunity to re-consider the LC potential for 
new benchmark scenarios with well-defined requirements in terms of
measurement accuracies, useful to optimize the detector design.

Within the constrained MSSM (cMSSM) parameter space, there are three regions, 
compatible with the LEP-2 constraints, where the lightest neutralino, $\chi$, 
provides a cold dark matter density $\Omega_{CDM} h^2$ matching the 
WMAP results. These are i) the co-annihilation ($\chi \tilde{\ell}$) region 
at small slepton-$\chi$ mass difference, the  rapid annihilation 
($\chi \chi \to A$) region at large values of $\tan \beta$ and the 
focus point region. Assessing the LC potential in testing the nature 
of dark matter requires to determine its reach and accuracy along these 
regions, while varying the reduced set of free parameters ($m_{1/2}$, $m_0$, 
$\tan \beta$, $A_0$) and study the experimental implications.
The accuracy of the WMAP data reduces the dimensionality of the 
cMSSM parameter space, introducing relations between $m_0$ and $m_{1/2}$, 
for each value of $\tan \beta$, which will be referred to as 
WMAP-lines~\cite{Battaglia:2003ab}.

This study represents a preliminary survey of the capabilities and 
experimental issues for slepton and Higgs analyses of benchmarks 
situated along the co-annihilation region and in the rapid annihilation funnel.
It considers centre-of-mass energies, $\sqrt{s}$, of 0.5~TeV and 1~TeV. 
Events have been generated with {\tt Pythia 6.205}+{\tt Isajet 7.67}, 
including bremsstrahlung effects, full detector simulation has been performed 
with the {\tt Brahms} program and fast simulation with 
a modified version of {\tt Simdet 4.0} and the {\tt JAS-3} LCD software.

%\begin{figure}
%\begin{center}
%{\epsfig{file=cdm_region2.eps,width=8.5cm}}
%\end{center}
%\caption[]{...}
%\label{fig:1}
%\end{figure}

\section{Slepton Reconstruction in co-Annihilation Tail}

The co-annihilation tail, at moderate $\tan \beta$ values, runs along the 
lower edge of the allowed cMSSM parameter space, up to $m_{1/2}$ values 
of $\simeq$~900~GeV. At the LC, the highest reach in $m_{1/2}$ comes from 
right-handed slepton pair production, 
$e^+e^- \to \tilde \ell^+_R \tilde \ell^-_R$. At 1~TeV, the LC sensitivity 
extends up to the extreme tip of the 
co-annihilation tail for $\tan \beta$ = 5 - 10, also beyond the LHC reach for 
sleptons, which stops at $m_{1/2} \simeq$~500~GeV. 
Along the WMAP line, $\tilde \ell_R$ 
becomes nearly degenerate with $chi$, resulting in soft leptons production. 
Tuning $\sqrt{s}$, to maximize the production cross section $\sigma$, further 
softens the lower lepton energy endpoint, $E_{\ell}^{min} = 
\frac{1}{2} M_{\tilde\ell} \left(1-\frac{M^2_{\tilde\chi^0_1}}
{M^2_{\tilde\ell}}\right) \gamma (1 - \sqrt{1-\frac{M^2_{\ell}}
{E^2_{beam}}})$. At $\tan \beta$=5, 10 $E_{\ell}^{min}$ is as low as 
1.5~GeV and 4~GeV respectively. In these scenarios, lepton identification 
becomes critical due to the intrinsic momentum cut-off of the ECAL and muon 
chambers and to the $\gamma \gamma \to {\mathrm{hadrons}}$ background. In the 
{\sc Tesla} detector design with $B$=4~T, the lower momentum cutoffs are 
1.5~GeV and 4.2~GeV respectively. The process 
$e^+e^- \to \tilde{\ell}^+_R \tilde{\ell}^-_R \to \ell^+ \chi^0_1 \ell^- 
\chi^0_1$ has been studied at 1~TeV for $m_{1/2}$= 600~GeV, 800~GeV and 
950~GeV with $\tan \beta$=5. The momentum acceptance for lepton identification
cuts into the lower momentum 
endpoint for $m_{1/2} \ge$~600~GeV, thus making difficult 
the extraction of the slepton-$\chi$ mass difference from the lepton momentum 
spectrum. This problem can be mitigated by using the specific ionization, 
$\frac{dE}{dx}$ in the Time Projection Chamber, to identify low momentum 
electrons. Assuming 200 samplings and 4.5~\% resolution, the $\frac{dE}{dx}$ 
provides $\ge 4 \sigma$ $e/\pi$ separation for $p > 0.9$~GeV. This recovers 
the accessibility of the lower energy endpoint for selectrons, and thus the 
$\tilde{e}$-$\chi$ mass difference. A fit to the reconstructed energy 
spectrum, using the full {\tt Geant-3} simulation, shows that the mass 
difference can be measured with a statistical accuracy of 0.02~\% to 
0.03~\% for 600~GeV $< m_{1/2} <$ 950~GeV.

\section{$\chi$ and $A^0$ Reconstruction in Rapid Annihilation Funnel}

In the rapid annihilation funnel, the dark matter density $\Omega_{CDM} h^2$ 
is controlled by the value of $R= 2 M_{\chi}/M_A$ and $\tan \beta$. 
A scan of $m_0$, $m_{1/2}$, $\tan \beta$, performed with 
{\tt microMEGAS}~\cite{Belanger:2004yn}, 
imposing $M_{h^0} > 112$~GeV and 0.093 $< \Omega_{CDM} h^2 <$ 0.129, shows 
that the derivative $\delta \Omega_{CDM}/\delta R$ is 2-5, depending on the 
exact position and the value of $\tan \beta$. Similar results have been 
obtained using {\tt DarkSUSY}~\cite{Gondolo:2004sc}. 
This indicates that the ratio of the $\chi$ 
to $A^0$ boson masses must be measured to better than 1~\% for predicting 
$\Omega_{CDM} h^2$ to an accuracy comparable to that expected by the 
next generation of satellite experiments. It is interesting to observe that, 
in the cosmologically favored funnel region, all the Higgs bosons are 
within the LHC sensitivity~\cite{Battaglia:2004mp}. However, 
the LHC accuracy on $M_{A^0}$ critically depends on the availability 
of the $A^0 \to \mu^+ \mu^-$ channel. This will be at the borderline of the 
LHC sensitivity for a significant part of CDM-favored region. Therefore, it 
is important to assess the LC potential in studying the $e^+e^- \to H^0A^0$ 
channel~\cite{Desch:2004yb}. 
A study point has been defined, corresponding to $m_0$=380.00~GeV, 
$m_{1/2}$=420.00~GeV, $\tan \beta$=53, $A$=0, $Sgn(\mu)$=+1 and 
$M_{top}$=178~GeV. These parameters give $M_{A^0}$=419~GeV, 
$M_{\chi^0_1}$=169~GeV and $M_{\tilde{\tau_1}}$=195~GeV. The analysis 
strategy at the LC is to determine first 
$M_{\tilde \tau_1}$ by threshold scan. A two-point scan at $\sqrt{s}$ 
values of 425~GeV and 500~GeV with 200~fb$^{-1}$ and 300~fb$^{-1}$ 
respectively, will provide an accuracy 
$\delta M_{\tilde \tau_1} / M_{\tilde \tau_1} \simeq 0.5~\%$.
Then $M_{\chi}$ can be extracted from the 
$\Delta M=M_{\tilde \tau_1}-M_{\chi}$ mass difference using the
$\tilde \tau_1 \to \tau \chi$ jet kinematics, at $\sqrt{s}$=0.5~TeV. 
The estimated accuracy is $\delta M_{\chi}/M_{\chi} \simeq$ 0.8~\% 
for 500~fb$^{-1}$ of data.
\begin{figure}[ht!]
\begin{center} 
\epsfig{file=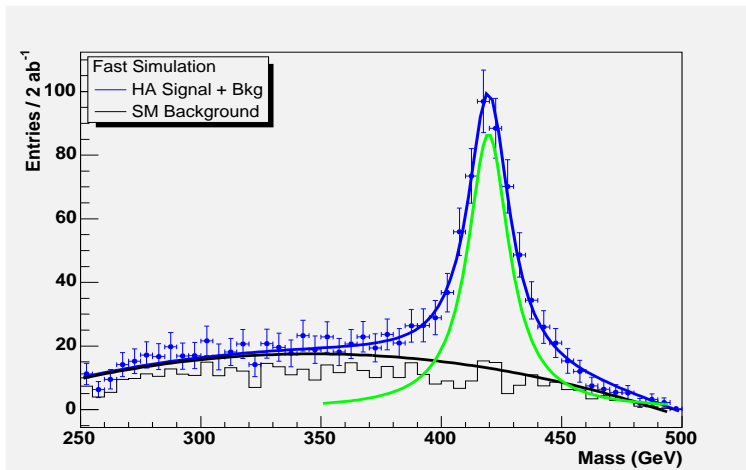,width=10.0cm,height=6.25cm}
\end{center}
\caption{Di-jet mass distribution for the $H^0A^0$ analysis. 
The histograms show the reconstruction 
results for the signal+background and the background alone and the 
curves the results of the 5-parameter fit.}
\label{fig:massfit}
\end{figure}
At 1~TeV, the effective 
$e^+e^- \to H^0 A^0 \to b \bar b b \bar b$ production cross 
section for the chosen benchmark point is 0.9~fb and the final state can be 
separated from both the $Z^0 Z^0$, $W^+ W^-$ and the inclusive 
$b \bar b \bar b$ backgrounds. After event selection, the di-jet pairing which
minimizes the di-jet mass difference has been chosen and the di-jet mass 
resolution improved by applying a 4-C fit. The resulting mass distribution 
is shown in Figure~\ref{fig:massfit}. 
The $A^0$ mass, $M_A$, and width, $\Gamma_A$ have been 
extracted by a multi-parameter fit leaving the parameters of the Breit-Wigner 
signal and second-order polynomial background free. The fit has been repeated 
by including also the $M_A - M_H$ mass splitting, or by constraining it to the 
model value. Results are summarized in Table~\ref{tab:massfit}.

\begin{table}[hb!]
\begin{center}
\begin{tabular}{|c|c|c|}
\hline
                & 6-par Fit       & 5-par Fit      \\ \hline
$M_A$ (GeV)     & 415.9$^{+2.5}_{-1.4}$ & 418.9$\pm$0.8 \\
$\Gamma_A$ (GeV) & 11.5$\pm$4.8 & 16.1$\pm$2.7 \\
$M_H-M_A$ (GeV) & 8.5$^{+2.3}_{-5.2}$ & 1.4 (Fixed) \\
\hline
\end{tabular}
\end{center}
\caption{Results of the fit for 2~ab$^{-1}$ of data at 1~TeV}
\label{tab:massfit}
\end{table}

\section{$\Omega_{CDM}$ Predictions from the LC Data}

In this study, the 
accuracy corresponding to the LC data precision has been evaluated, within 
the cMSSM, for the fast annihilation funnel benchmark point. 
\begin{figure}[hb!]
\begin{center} 
\epsfig{file=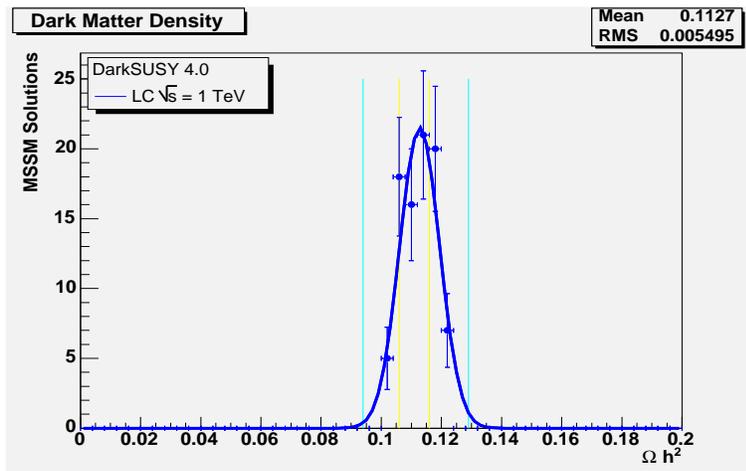,width=10.0cm,height=6.25cm}
\end{center}
\caption{Probability density function of $\Omega_{CDM} h^2$ for the 
cMSSM solutions compatible with the LC measurements of $M_A$, $M_{\chi}$ and 
$M_{\tilde \tau_1}$. The vertical lines show the uncertainty of the WMAP 
result (light blue) and that expected by PLANCK (yellow).}
\label{fig:omega}
\end{figure}
First the dependencies of $\Omega_{CDM} h^2$ on SUSY and SM parameters have
been evaluated. Then a scan 
of the cMSSM parameter space has been performed using {\tt DarkSUSY 4.0}.
The constraints of the masses obtained in the previous section have been 
included together with $\tan \beta \pm 3$, $M_{top} \pm 0.1$~GeV and 
$M_b \pm 0.05$~GeV. Large SUSY corrections to $\Gamma_A$ also affect the 
computations of $\Omega_{CDM} h^2$. These can be controlled both by the direct 
determination of the $A^0$ boson width and by that of the 
$h^0 \to b \bar b$ branching fraction. In fact 
${\mathrm{BR}}(h^0 \to b \bar {b})$ is sensitive to the same $\Delta m_b$ 
shift of the $b$-Yukawa coupling, due to SUSY loops, responsible for the 
corrections to $\Gamma_{A^0}$. The precision of the LC data can control the 
$A^0$ width to 10~\% by relating 
${\mathrm{BR}}(h^0 \to b \bar{b})$ to $\Gamma_{A^0}$ 
as: $\Gamma_{A^0} = \frac{{\mathrm{BR}}(h^0 \to b \bar {b})}
{{\mathrm{BR}}(A^0 \to b \bar {b})} \times \Gamma_{h^0} \times \tan^2 \beta$.
A direct scan at a $\gamma \gamma$ collider may improve this accuracy. 
The resulting p.d.f. for $\Omega_{CDM} h^2$ is 
shown in Figure~\ref{fig:omega}. 
The relative accuracy is 
$\frac{\delta \Omega h^2}{\Omega h^2} = 
\pm 0.049~{\mathrm{(SUSY~Masses)}}
\pm 0.050~{\mathrm{(SUSY~Corr.)}}
\pm 0.035~(M_{{\mathrm{quarks}}})$.
 
\section{Conclusion}

The accuracy in the measurement of the masses of sleptons and heavy Higgs 
bosons in cMSSM scenarios compatible with the WMAP results on cold dark matter,
has been re-analyzed in view of the requirements for predicting $\Omega h^2$ 
to a few~\% from SUSY measurements. Results show that the typical 
${\cal{O}}(0.1~\%)$ accuracy on slepton masses is realisable along 
the co-annihilation tail. At small $M_{\tilde \ell} - M_{\chi^0_1}$ values, 
this region is characterized by momenta of the emitted leptons which require 
to extend the acceptance of lepton identification and a careful study of 
the $\gamma \gamma$ background rejection.
The $A$~funnel region presents an interesting analysis program at 0.5~TeV and 
1.0~TeV, where large data sets should provide a $\Omega_{CDM} h^2$ accuracy 
of ${\cal{O}}(5~\%)$ and control of systematics to a comparable level.

\end{document}